# Magnetic Field Gated and Current Controlled Spintronic Mem-transistor Neuron -based Spiking Neural Networks

Aijaz H. Lone ,*,† Meng Tang,‡ Daniel N. Rahimi,† Xuecui Zou,† Dongxing Zheng,‡ Hossein Fariborzi,† Xixiang Zhang,‡ and Gianluca Setti*,†

†*Computer, Electrical and Mathematical Sciences and Engineering (CEMSE) Division,King Abdullah University of Science and Technology (KAUST), Saudi Arabia*

‡*Physical Sciences and Engineering (PSE) Division, King Abdullah University of Scienceand Technology (KAUST), Saudi Arabia*

E-mail: aijaz.lone@kaust.edu.sa; gianluca.setti@kaust.edu.sa

## Abstract

Spintronic devices, such as the domain walls and skyrmions, have shown significant potential for applications in energy-efficient data storage and beyond CMOScomputing architectures. In recent years, spiking neural networks have shown more bio-plausibility. Based on the magnetic multilayer spintronic devices, we demonstrate the magnetic field-gated Leaky integrate and fire neuron characteristics for the spiking neural network applications. The LIF characteristics are controlled by the current pulses, which drive the domain wall/bubble motion, and an external magnetic field is used as the bias to tune the firing properties of the neuron. Thus, the device works like a gate-controlled LIF neuron, acting like a spintronic Mem-Transistor device. We develop a LIF neuron model based on the measured characteristics to show the device's integration in the system-level SNNs. We extend the study and propose a scaled version of the demonstrated device with a multilayer spintronic domain wall magnetic tunnel junction as a LIF neuron. using the combination of SOT and the variation of the demagnetization energy across the thin film, the modified leaky integrate and fire LIF neuron characteristics are realized in the proposed devices. The neuron device characteristics are modeled as the modified LIF neuron model. Finally, we integrate the measured and simulated neuron models in the 3-layer spiking neural network(SNN) and convolutional spiking neural network CSNN framework to test these spiking neuron models for classification of the MNIST and FMNIST datasets. In both architectures, the network achieves classification accuracy above 96%. Considering the good system-level performance, mem-transistor properties, and promise for scalability. The presented devices show an excellent alternative for neuromorphic computing applications.

*keywords: Spintronics, Spintronic mem-transistor, Domain Walls, LIF neurons, MTJ, SNN, CSNN, Neuromorphic devices,*

## Introduction

Spintronics has shown a promising future for data storage and computing. [1,2] In particu- lar, owing to tuneable physical properties in spintronic devices, diverse functionalities have been explored. Depending upon the different device characteristics, such as non-volatile bi-nary and analog switching, stochastic switching. Spintronics has shown remarkable presence in unconventional computing architectures like neuromorphic computing, [3–5] probabilistic computing, [6–8] and reservoir computing paradigms. [9,10] In particular, over the past decade, spintronic devices based on the domain walls, vortices, and skyrmion dynamics have been extensively pursued for applications in data storage [11] and computation. [12,13] The DWs and skyrmions are experimentally stabilized in a ferromagnetic thin film system when different magnetic energy terms such as magnetic anisotropy, exchange energy, stray field energy and Dzyaloshinskii-Moriya interaction (DMI) compete with each other to attain the energy min- ima. This results in a domain with all spins in the $+z$ axis and another domain in the $-z$ axis, where the two domains are connected by a region in which spins rotate from $+z$ to $-z$ in a continuous fashion. [11,12] The stabilization of stripe-like domain walls has been shown in the ferromagnetic multilayer systems such as $(Co/Pt)_n$, $(CoFeB/Pt)_n$ and $(CoFeB/Ta)_n$. [14,15] These domain walls are driven by spin transfer torque STT, SOT, or external magnetic field. The movement of the domain wall creates variations in the net magnetization of the MTJ-free layer, creating an analog MTJ resistance switching characterized by the tunneling magnetoresistance TMR. [16] Magnetic Skyrmions can be considered circular domain walls formed after adding the chiral asymmetric energy Dzyaloshinskii-Moriya interaction (DMI), which emerges at the heavy metal/ferromagnet thin film interface with strong Rashba spin- orbit coupling. The domain walls and skyrmions have emerged as promising alternatives for racetrack memory, neuromorphic computing, [17–21] and reservoir computing. [10] Especially in neuromorphic computing technologies, the magnetic domain wall and skyrmion-based mag- netic tunnel junction structures have been proposed to demonstrate the synaptic and neuron behaviors in artificial neural network architectures. [22–25] More recently, significant research has been spurred into spiking neural networks SNNs, which are considered to have more bio-logical plausibility. Spiking Neural Networks (SNNs) are computational models that process information based on the number of spikes generated by the spiking neurons. Unlike traditional artificial neural networks that process information continuously, SNNs operate using discrete time events or "spikes.[26,27] SNNs utilize spatiotemporal data encoding, giving them more energy efficiency and high performance for event-driven AI on edge computing [28–34]tasks. Some spiking neurons have recently been proposed based on different spintronic de-vices. However, most of these proposals are restricted to the simulations only,[35–37] apart from a few experimental works, [38,39] which have shown spiking neurons -based on magnetization oscillations. But, to our knowledge, the practical realization of a leaky integrated fire (LIF) neuron is yet to be available. This paper experimentally demonstrates the practical realization of multilayer ferromagnetic spintronic devices exhibiting biologically inspired leaky integration and fire neuron (LIF) behavior. These LIF neuron devices are based on the magnetic domain wall and/or skyrmion switching in the [Ta/CoFeB/MgO]x20 ferromagnetic multilayer system. The current pulses applied laterally drive the domain walls, resulting in gradual switching. The magnetic field bias is used to tune the firing characteristics of LIF neurons. We also propose a scalable micromagnetic version of the LIF device. The characteristics of the experimental/simulated neuron device are modeled as the modified LIF neuron. The models are integrated into the 3-layer spiking neural network (SNN) and convolutional CSNN framework to test these spiking neurons for classifying the MNIST and FMNIST datasets. In both architectures, the network achieves classification accuracy above 96%. Additionally, the LIF neuron latency is in ns; thus, when integrated with the CMOS, the proposed device structures and associated systems exhibit an excellent future for energy-efficient neuromorphic computing.

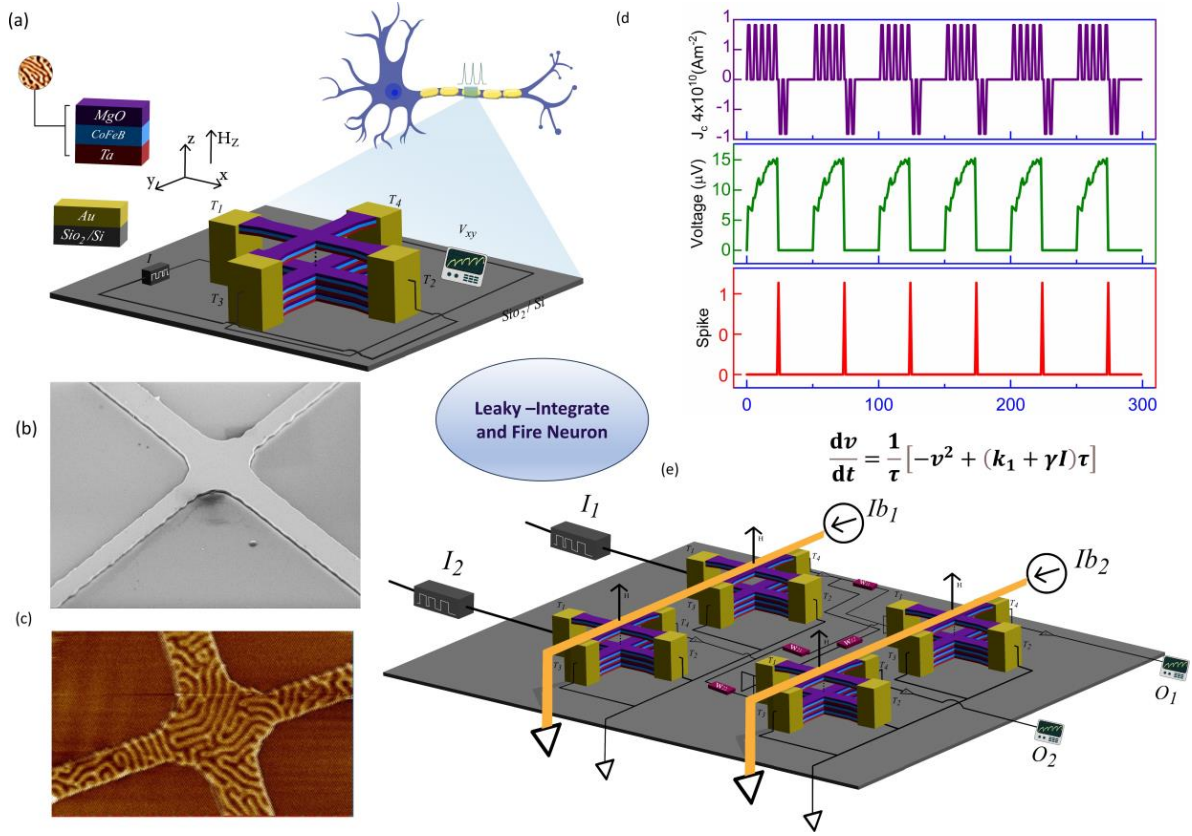

**Figure 1.** (a) Domain wall crossbar devices. (b) Equivalent resistor model.

## Magnetic multilayer LIF Neuron Device Structure

The multilayer spintronic LIF neuron device structure is shown in Fig. 1(a). The device structure comprises the (Ta/CoFeB/MgO)x20 multilayer system. Devices were fabricated as crossbars with asymmetry such that one arm is $(3 \times 50 \, \mu m^2)$, while the other is $(2 \times 50 \, \mu m)$, as shown in the scanning electron microscopy SEM image in Fig. 1(b). At zero field, we observe stabilization of the narrow stripe domains in the device as shown in the magnetic force microscopy MFM image (see Fig. 1(c)). In the Ta/CoFeB multilayer systems, the chiral Neel-type stripe DWs are stabilized due to DMI, and the direction of motion depends on the sign of DMI. The current pulses are applied between terminals T-1 and T-2, which generate the spin transfer torque STT, joule heating, and small SOT in the CoFeB thin film. The expected mild joule heating in the sample assists the STT and SOT in gradually switching the magnetization of the device by magnetic field-assisted current-controlled domain wall motion. The SOT emerges from the Rashba spin-orbit coupling at the Ta/CoFeB interface and bulk giant spin Hall effect in the broken inversion symmetry systems. The STT andSOT act on the DW and/or skyrmion spins, transferring angular momentum adiabatically and non-adiabatically, resulting in DW motion. We apply a perpendicular magnetic fieldto bias the magnetization of the device at a desired resistance value, which we categorize into two regions, namely excitatory and inhibitory regions. Then, using continuous current pulses ranging from $2 \times 10^{10}$ Am$^{-2}$ $\mu$s to $7 \times 10^{10}$ Am$^{-2}$ $\mu$s , the magnetization switchingis achieved to realize integrate and leak behavior. Assuming a certain voltage threshold for firing (determined by the read OP-AM Comparator), we reset the device to its rest potential by a magnetic field. The integrate and leak characteristics are obtained by employing the three significant effects: (1) Domain wall pinning/depinning, (2) magnetic field-assisted current-driven domain wall dynamics, (3) thermal effects, and (4) fabrication asymmetries in the crossbar dimensions. The observed characteristics mimic the typical leaky-integrate and fire (LIF) neuron, considered one of the most biologically plausible neuron models in neuroscience and spiking neural networks SNNs. Especially the leakage in the neuron plays a crucial role in encoding the timing information, which is essential to differentiate the old and new input spikes. The anomalous Hall voltage directly depends on the magnetization switching.

$$\rho_{xy} = \rho^O + \rho^A + \rho^T = R_O B + R_S \mu_0 M_Z + R_O P \underset{em}{B}^z \tag{1}$$

Fig. 1(d) shows the leaky integrate and fire neuron characteristics realized by applying the current pulses with density $J_c = 6 \times 10^{10}$ A/m². and biasing the device at 220Oe external magnetic field. The current pulses of pulse width 500 $\mu$s assist and magnetic field assist each other in gradually switching the magnetization by combining spin transfer torque **STT**, thermal effects, magnetic field driven DW motion, and possible spin-orbit torque **SOT** can't be neglected. At $J_c = 6 \times 10^{10}$ A/m²., which is 13mA current and field 220Oe, the voltage signals vary up to 17 $\mu$V. The voltage increases with the current pulses, and in the absence of the external current, we observe the leaky behavior, as shown in Fig.1(d). The read signals are compared to the fixed threshold voltage, 15 $\mu$V in this case, and a spike is generated as the voltage reaches 15 $\mu$V, as shown in Fig.1(d). We furthermore model this behavior using the differential equation presented in Fig.1. To utilize the magnetic field gating at the circuit level, we propose the schematic shown in Fig.1(e) here, the external magnetic bias can be achieved by passing a current via the wire (golden ribbon). When writing a specific device in the crossbar, the magnetic field is switched on first by passing the current Ib in the respective wire, followed by a continuous application of current pulses $I_1$, as shown in Fig.1(e). The output voltage is passed to a voltage comparator, which generates spikes propagating to the next layer weighted by the synapse $W_{ij}$. Fig. 2 explains the magnetic field gating for tuning the operating regime of the **LIF** neurons. The devices can be operated in 2 regions depending on the magnitude of the external magnetic field, as shown in Fig. 2(a). If the magnetic field is in the range of +150Oe to 230Oe or if the magnetic field range is (-230O to -150Oe), we observe evident hysteresis with a memory window. The observed output voltage characteristics are strong, as shown in the Fig.2(b) heat-map. The solid red signals indicate output voltage in the range of (20 to 30)$\mu$V. We call the neurons working in this region the excitatory neurons. These neurons have a very strong spiking probability when stressed with current/voltage pulses. In the case of a biasing field in the range (-100 to 100)Oe or above/below 250Oe/-250Oe, the measured output voltage signals are weak, indicating that the neurons working in this region have very low spiking probability, as shown in Fig.2(c).In the case of the low magnetic field and in the saturation range, the hysteresis loop is pinched, which shows no memory effect.

Thus, the probability of firing these neurons is low. The magnetic field-gated current-controlled output voltage results demonstrate that the device behaves like a spintronic transistor, a (memory transistor) or a mem-transistor. This plays a vital role in the circuit-level implementations where the crossbars -based on these devices can be realized. The need for the access transistor in the memory crossbar array can be avoided using this concept of magnetic field-assisted current-driven spintronic LIF neurons. We provide a DC across its field wire to access the specific device. Then, the devices are excited by the low current (4 $\times 10^{10}$ A/m$^2$) pulses. The current will not affect its firing probability unless the device is in an excitatory region. That mitigates the writing error problem faced by the devices in a pseudo-crossbar array. The concept is further demonstrated in Fig.2(c), for the writing current equal to 15mA ($6.5\times10^{10}$ Am$^{-2}$) and varying magnetic biasing field. We observe that the output leaky-integrated voltage signals are maximum (20 $\mu$V) for H = 22OOe and ($-30\,\mu$V) for -22OOe. Whereas, for H=0Oe, 50OOe, and -50OOe, the voltage is almost 0, indicating potent inhibition after reaching the threshold voltage, the resetting of the neuron to a resting potential is shown. Fig.2(d) shows the LIF characteristics for the writing current equal to 10mA ($5 \times 10^{10}$ Am$^{-2}$), and biasing varied from -50OOe to 50OOe. Like in the previous case, we observe an excellent gating effect and current controlled LIF characteristics. Note: Some randomness in the resting potential, which originates from the domain pinning/depinning, is observed. In circuit implementation, mild randomness in the resting potential should not create any substantial error unless the resting voltage is near the threshold. The output spikes can't be generated. Fig. 3 shows the magnetic field assisted LIF neuron characteristics for different writing currents. Fig.3(a) shows the typical LIF behavior at writing current 6mA ($2.5 \times 10^{10}$ Am$^{-2}$) and magnetic field (-16OOe, -18OOe, -20OOe, -22OOe and -50OOe). The output voltage signal amplitude is equally reduced for reduced current density. The LIF signal increases as the field increases from 0Oe towards the center of the excitatory region where dV is 5 $\mu$V at -18OOe. As the device is biased away from the optimum excitatory region on either side, the output signal amplitude is reduced due to the limited memory window left for current-driven switching. The output first increases as we move away from the inhibitory region (pinched loop/small hysteresis) more toward the excitatory region (strong hysteresis). As we move away from the center of the excitatory region, the signal begins to drop as expected due to limited magnetization/domains, which the current can switch. The magnetic field gating is strong and independent of the writing current amplitude, and the quality of output LIF voltage is improved due to lowered random pinning/depinning as expected at low current densities. These results show the robustness of the magnetic field-assisted current-controlled concept. In Fig. 3(b), we show the measured LIF behavior at fixed bias 200Oe and writing current 6mA, 8mA, 10mA, and 15mA. We observe the strongest voltage signal of amplitude around 22 $\mu$V for 15mA. The output voltage signal amplitude is reduced as the writing current is lowered. Furthermore, the integration and leakage behavior becomes more stable at low currents due to smooth domain wall pinning/depinning at low writing current compared to abrupt pinning/depinning at 15mA. Fig. 3(c-f) shows the output signals at different magnetic biases and current pulses in the heat-map form. It can be seen clearly that best regions for excitatory device operations are obtained at writing current 15mA and magnetic field bias range (180Oe to 280Oe) [positive excitatory region-red] and (-200Oe to -320Oe) [negative excitatory region-blue]. As we lower the current, the output voltage starts diminishing, and the range of possible bias points is also reduced, as seen in Fig. 3(d-f). Compared to the 15mA write current, the signals at 6mA are almost negligible, as shown in Fig. 3(f). The magnetic field and write current dependence of the device output demonstrate a versatile neuron operation, as discussed in Fig. 2 and Fig. 3. By having this transistor-like capability where the magnetic field gating works like the gate voltage in a transistor and the write current between T-1 and T-2 acts like source-drain voltage. Thus, the presented device promises scope for integrability in a large neuromorphic system.

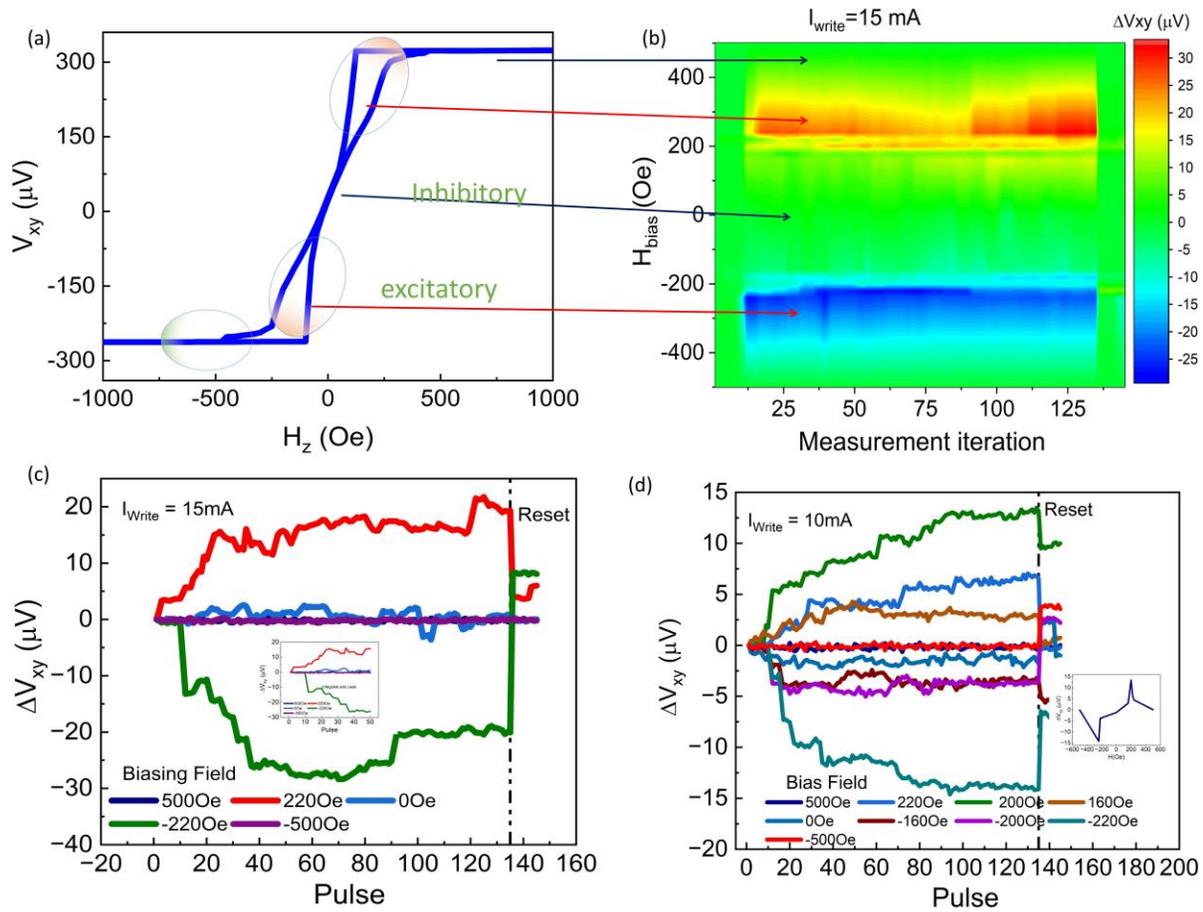

**Figure 2.** (a) Resistance- magnetic field hysteresis characteristics showing the regions of device operation. (b) LIF characteristics, such as the function of the magnetic field and the number of current pulses (measurement iteration), show the excitatory and inhibitory regions of operation. (c) current driven field gated
- LIF behavior for 15mA write current and magnetic field (-500Oe, -220Oe, 0, 220Oe and 500Oe). (D) current driven field gated - LIF behavior for 10mA write current and magnetic field (-500Oe, -220Oe, -200Oe, -160Oe, 0, 160e, 200Oe, 220Oe and 500Oe).

In any spiking neural network, the information is encoded majorly in neurons spiking frequency. To demonstrate the significance of magnetic field biasing and writing current on the spiking property of the LIF neurons. Fig. 4(a) shows the LIF behavior executed 100 times by the device in a continuous measurement setting. We applied 10 writing current pulses for 1 LIF operation; the neuron state is read 4 times between two consecutive writing pulses. Once the neuron voltage reaches a certain peak (threshold), we reset the neuron using an external magnetic field. The process was repeated 100x to estimate the mean peak voltage and the variance due to random domain wall pinning/depinning, as shown in Fig. 4(a). We observed a reliable neuron integration and fire followed by reset behavior throughout the 100 cycles.

Fig. 4(b) shows the LIF device characteristics at the bias field: 180Oe, 200Oe, 220Oe, and 240Oe; the device is set-reset 10 times. The neuron is reset back to a resting voltage in the proximity of 0V approximately. In the repeated setting and resetting of the device, we observe fluctuations in peak voltage, which is attributed to the stochastic domain pinning/depinning in a large-area device. We expect this stochasticity to be minimized in smaller-dimension devices. The variations in peak voltage and resting voltages for different biasing points are quantified statistically, as shown in Fig. 4(c-e). Fig. 4(c) shows the peak voltage distribution at bias field 180Oe; the mean peak voltage is $215\mu$V with a slight standard deviation of about $2.6\mu$V. Likewise, for 200Oe, the mean peak voltage is $238\mu$V, and the deviation is $4.65\mu$V. For 220Oe, the men are $266.1\mu$V, and the standard deviation is $7.8\mu$V. These results indicate that the device performance is optimum at a biasing field equal to 180Oe. We define a range of thresholds for the actual circuit realization to overcome the mild deviations in the peak voltages. A hysteretic comparator working in the respective range will generate a spike whenever the device voltage lies in the threshold voltage range. Likewise, the device is assumed to be in complete rest if the read voltage is in the resting range. In the case of ideal reset conditions, we measure integrated leaky behavior from the device and post-process the data in ex-situ conditions. We set a voltage threshold as a comparator, which generates spikes. As shown in Fig. 4(f), for the magnetic field 180Oe, 200Oe, and 220Oe, the threshold voltage is set at $17\mu V$. With the application of the writing current pulses, the voltage increases in LI behavior. As the voltage reaches the $V_{TH}$ range $V_{TH} = 17+-2.6~\mu V$ t, the comparator (simulation) generates an output spike, and at the same time, the devices are reset to the initial position. Likewise, for 200Oe and 220Oe the threshold voltage ranges are ( $V_{TH} = 28+-4.65~\mu V$ and $V_{TH} = 37+-7.8~\mu V$

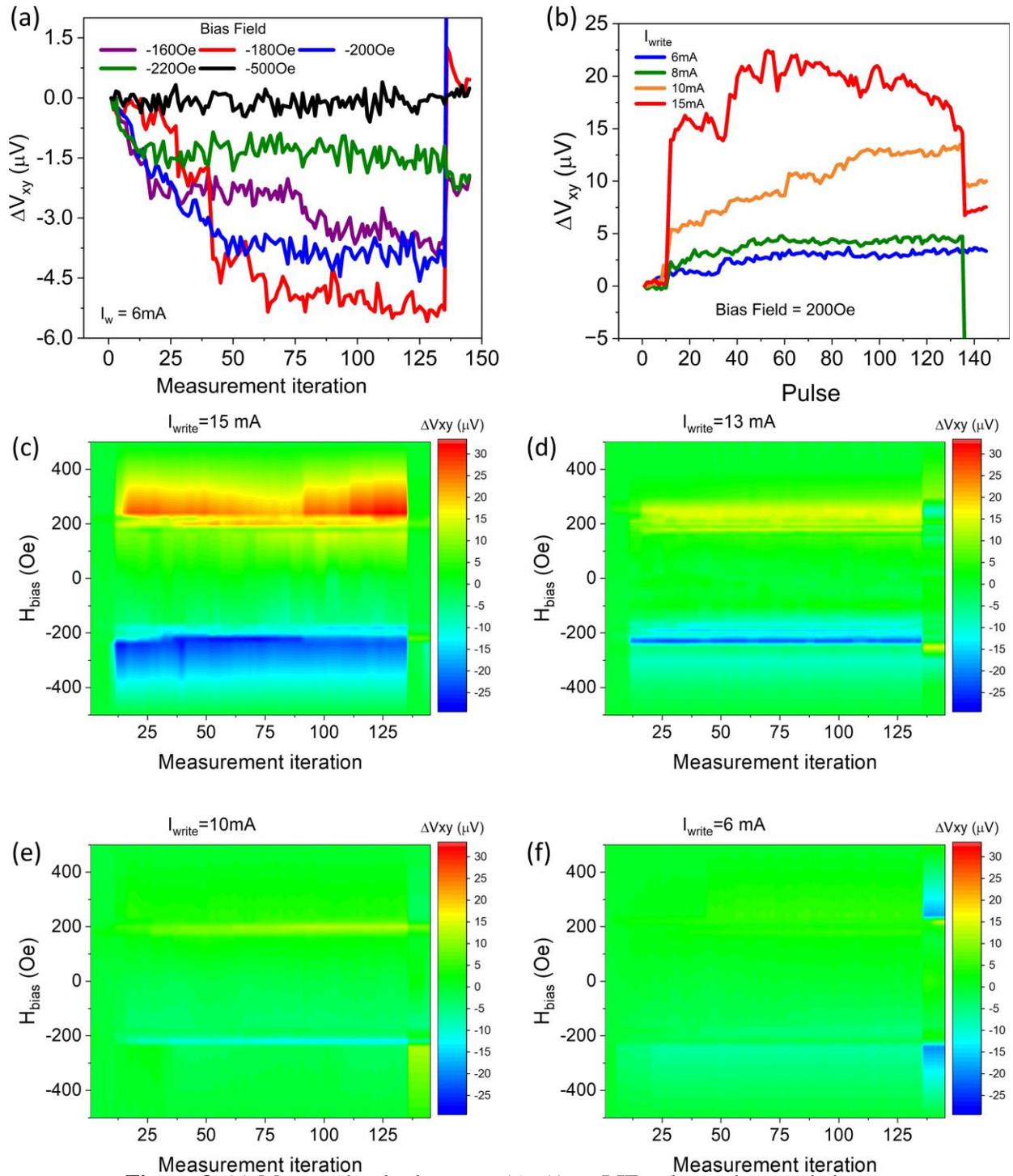

**Figure 3.** (a) Measured pulsed current (6mA) vs LIF voltage characteristics at different fields. (b) LIF voltage characteristics at constant field 200Oe and write current equal to 6mA, 8mA, 10mA, and 15mA. LIF voltage signal as a function of the magnetic field and current pulses (measurement iteration) for different write current amplitude. (c)15mA, (d) 13mA, (e) 10mA, and (f) 6mA.

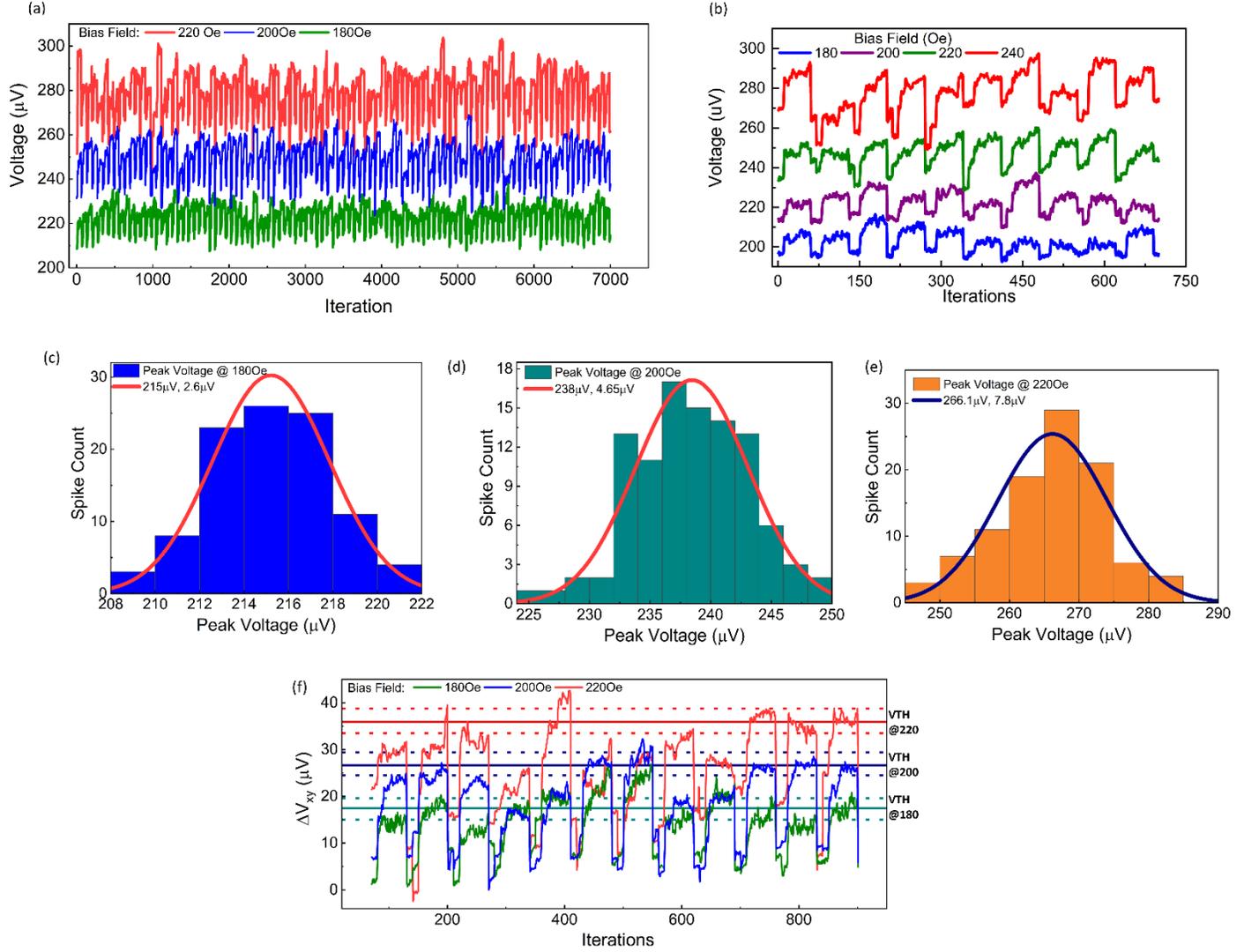

**Figure 4.** (a) (a) In-situ LIF characteristics for bias field 180Oe, 200Oe, and 220Oe repeated 100 times. (b) In-situ LIF characteristics for bias field 180Oe, 200Oe, 220Oe, and 240Oe repeated 10 times. Peak voltage distribution (c) 180Oe [$\bar{v} = 215$μV, $\sigma = 2.6$μV] (d) 200Oe [$\bar{v} = 238$μV, $\sigma = 4.65$μV]and (e) 220Oe [$\bar{v} = 266.1$μV, $\sigma = 7.8$μV]. (f) Threshold voltage ranges from 180Oe, 200Oe, and 220Oe.

So, in the presented devices, the magnetic fieldand the writing current amplitude can control the spiking rate. This increases the device's versatility and promises its significance in hardware SNN implementations.

# Multilayer Domain Wall LIF Neurons (Micromagnetics)

To study the viability of the multilayer spintronic devices as LIF neurons in the spiking neural network, scalability and increased output signal are highly desired. We propose the ($256 \times 64$ nm$^2$) spintronic LIF neuron device structure using micromagnetic simulations. The device structure consists of the (Ta/CoFeB)$_{x20}$ multilayer nanotracks in which either DWs or skyrmions can be moved from one end ($T_1$) to the other ($T_2$). The STT and SOT generated from the laterally flowing charge current drive the DW. The writing is carriedout by applying current pulses of amplitude $1 \times 10^{11}$ A/m$^2$ and width 1ns. To study the frequency response, period T varies from 1.5ns to 4ns. The neuron membrane potential (magnetization) reading is carried out by the tunneling magnetoresistance TMR effect viaan MTJ reading block, as shown in Fig. 1(a). As the DW moves across the free layer, itsees edge forces, and as it approaches the $T_2$ end of the device, the demagnetization energyincreases, opposing the DW movement. By reading magnetization using MTJ only nearthe $T_2$ end of the nanotrack, the integration of the magnetization in the presence of the current is realized, followed by the magnetization leakage in the absence of the current. The MTJ output terminal $T_3$ is connected to the comparator to generate the output spikes that propagate to the next layer. We use a negative resetting current pulse to reset the membrane potential. As the membrane potential reaches the threshold voltage $V_{th}$, thecomparator generates spikes and, at the same time, activates the resetting MOSFET, whichallows a resetting current pulse in the opposite direction. This current pulse forces the DW back to its original state, and membrane potential is reset to resting potential Vrest.

In Fig. 3, we show the different electrical and magnetic characteristics of the DW LIF neuron device. Fig. 3(a) shows how the demagnetization energy evolves as the DW moves towards the terminal $T_2$. We initiate the DWs at -200nm from the center towards the left.

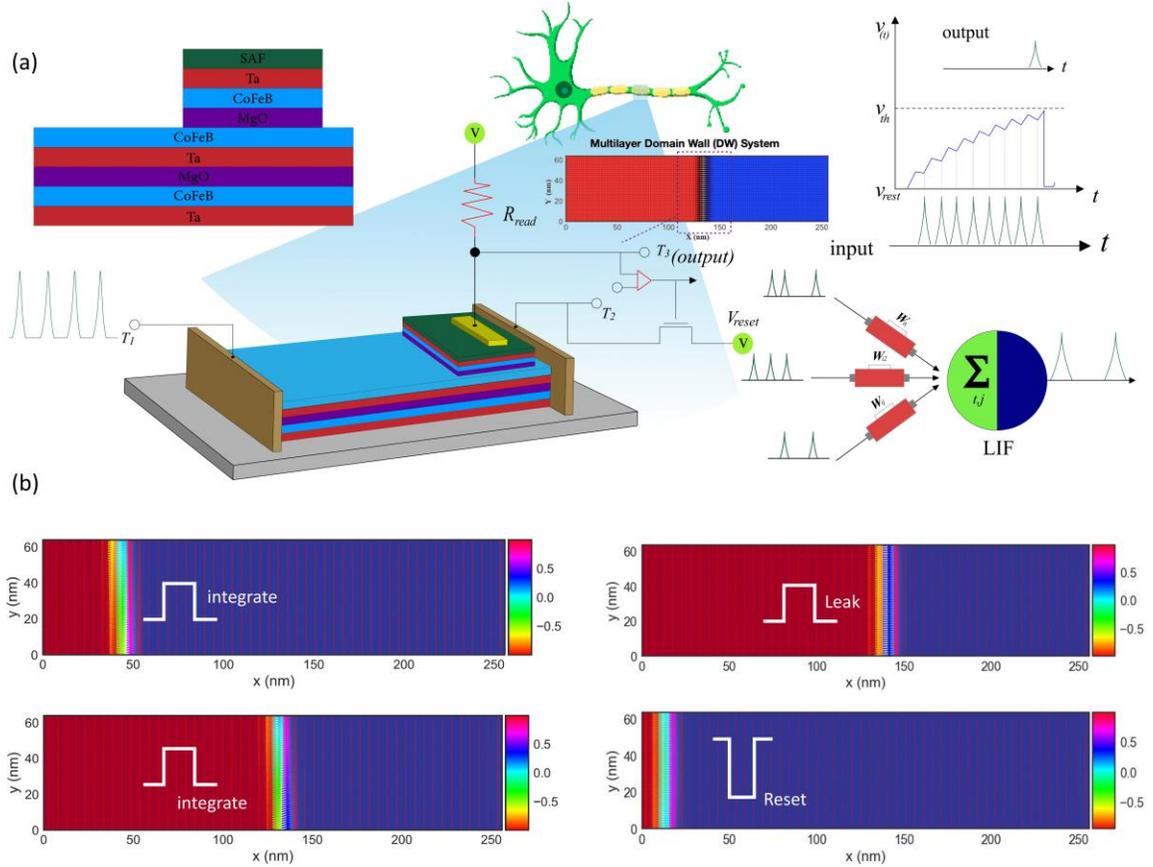

**Figure 5.** (a) Multilayer Spintronic LIF Magnetic tunnel junction device (Micromagnetics). (b) Magnetization evolution throughout the LIF process.

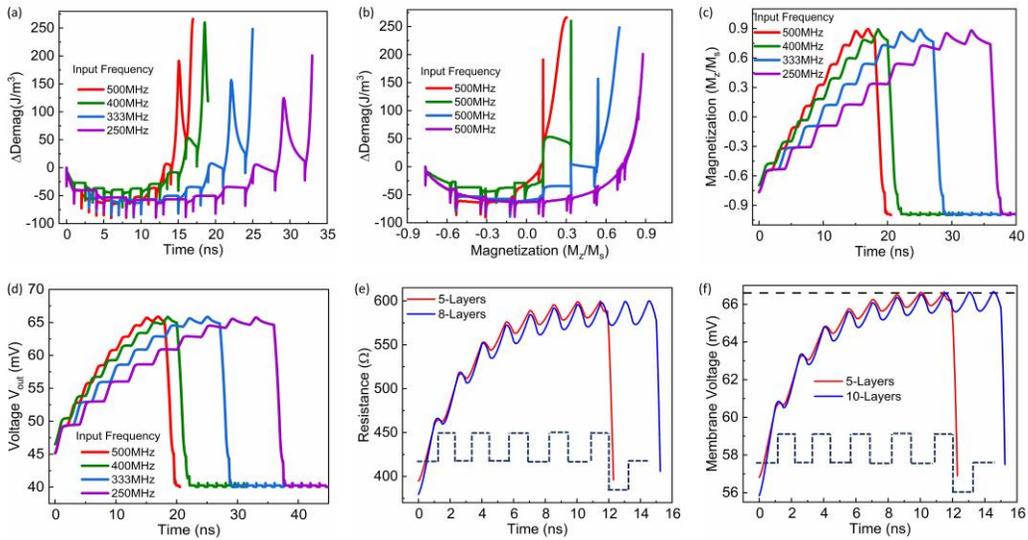

**Figure 6.** (a) demagnetization energy as function evolution with time. (b) Demagnetization energy as a function of domain wall position (in terms of magnetization). (c) Free layer magnetization at input frequencies (500MHz, 400MHz, 333MHz, and 250MHz) shows LIF characteristics. (d) Multilayer DW-LIF neuron output voltage at 500MHz, 400MHz, 333MHz, and 250MHz.(e) DW-LIF neuron tunnel resistance for 5 and 8-layer devices. (f) DW-LIF neuron output voltagefor 5 and 8-layer devices.

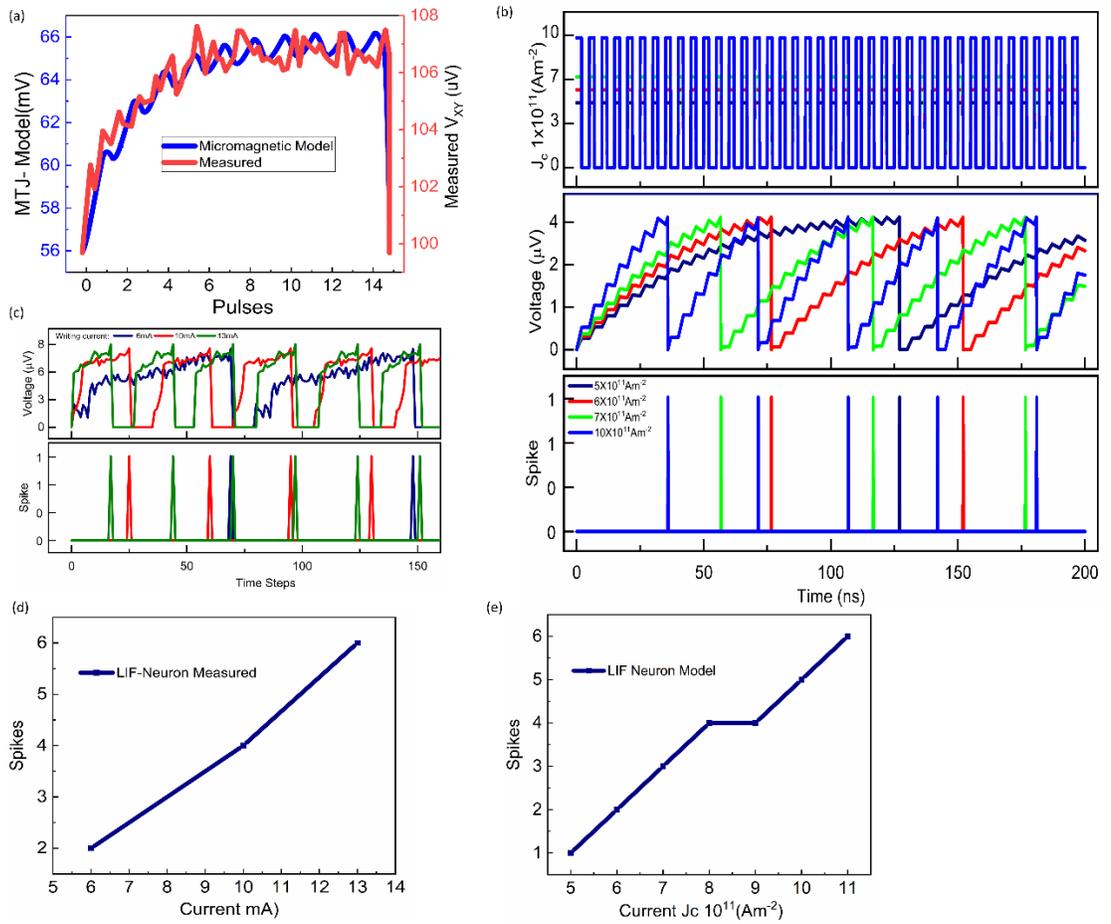

**Figure 7.** (a) Measured LIF device characteristics and the model. (b) LIF response of the model to increasing current densities shows increasing spiking frequency with increasing current density. (c) Ex-situ LIF characteristics at increasing current amplitudes (6mA, 10mA, and 13mA). (c) Increasing spike frequency with increasing current (measured). (e) Model-based results show increasing spike frequency with increasing current (simulation).

Thus, the multilayer thin film is magnetized 70% down (-z direction) and about 30% up (+z direction). The demagnetization energy is high, but as the DW starts moving right towards the center, the demagnetization is reduced until the DW reaches the center where domain sizes are equal; thus, demagnetization is minimal. The demagnetization is increased further as the DW continues its motion towards the terminal $T_2$. Fig. 6(b) shows the change in demagnetization energy for the magnetization of the multilayer. We observe the minima around $Mz = 0$, which corresponds to the DW at the center. Fig. 6(a-b) shows the increasing demagnetization for time/magnetization for 4 different input pulse/spike frequencies. We observe two components in the demagnetization time characteristics: (1) the application of current increases the demagnetization due to the SOT on the DW spins, which are forced from the stable in-plane to a perpendicular direction momentarily. When the current switches back to zero, the demagnetization also relaxes, but due to DW motion, the ratio of +z/-z domain spins changes. This gives rise to a shift in the demagnetization curve. Thus, we observe (1) a Transient volatile memory effect due to SOT and (2) a Stable non-volatile memory effect due to asymmetry in domain sizes. For a 50% duty cycle, a 1ns ON current followed by a 1ns OFF current pulses are applied, which means (500MHz) input frequency; likewise, for a 40% duty cycle, the ON current pulse width is 1ns followed by 1.5ns OFF state(400MHz) input frequency, for a 33% duty cycle (1ns ON/2ns OFF), the input frequency is 333MHz and for 250MHz(the pulse scheme is 1ns ON/3ns OFF). The DW reaches the threshold position in 16ns, and we observe no further increment in the demagnetization energy. For 333MHz, the threshold location is achieved in around 27ns. The demagnetization characteristics directly affect the overall motion of the DW. As the DW starts approaching terminal-$T_2$, it begins to see an increasing demagnetization force acting on it. The DW moves right in the presence of the SOT pulse, as shown in Fig. 5(b), but when the current is switched off, the demagnetization force starts dominating the overall motion; thus, we observe the DW moving back, as seen in Fig.5(b). The SOT effect is equivalent to integrating the membrane potential in the biological neuron. Meanwhile, the DW's relaxing back mimics the membrane potential leakage. As explained in the previous section, the DW motion changes the net magnetization of the MTJ-free layer. In Fig.6(c), the magnetization-time characteristics clearly show the integrate and leak behavior. For high input frequency (50% duty cycle), the normalized magnetization (Mz/Ms) reaches the threshold value of 0.9 much earlier when compared to the other two cases. In the case of lower frequencies, such as the 33% duty cycle, the DW gets enough time to relax; thus, magnetization leakage is increased, reflected in the blue curve. The MTJ reads the magnetization change in terms of its tunnel magnetoresistance. The magnetization is mapped to the MTJ resistance or neuron resistance by:

$$(2)$$

$$R_{\text{neuron}} = R_{\text{AP}} \ \frac{[1 - \overline{m} \cdot \overline{m} P]}{2} + R_P \ \frac{[1 + \overline{m} \cdot \overline{m} P]}{2}$$

Here, the $R_P/R_{AP}$ represents the MTJ resistance in a parallel/anti-parallel state. Fig.6(e) shows the neuron resistance switching for 2 devices with five and eight CoFeB layers. We considered the TMR 200% and Rp/RAP= 200/600. If the complete magnetization switching is considered, the out switches in the range (55 to 70)mV are shown in Fig. 6(f). The resistance increases from 380 $\Omega$ to 590 $\Omega$ in a leaky integrated behavior. For the $L_5$ case, we observe fast DW switching, whereas, for the $L_8$ case, the DW velocity is reduced. Thus, the threshold voltage is achieved in 15ns. Also, for increased CoFeB layers, the leakageconductance is increased due to increased demagnetization energy, as shown by the 8-layer resistance and MTJ neuron voltage characteristics in Fig. 6(d-e)—the MTJ $\mathbf{T_3}$.

$$V_{\text{out}} = \frac{R_{neuron}}{R_{neuron} + R_{read}} V_{\text{read}}, \qquad V_{\text{out}} \geq V_{\text{th}}: \text{spike} = 1 \qquad (3)$$
$$\text{Else}: \text{spike} = 0$$

The output of the MTJ neuron (terminal $T_3$) is connected **$T_3$** to a voltage comparator having threshold voltage $V_{th}$. As the MTJ neuron voltage reaches this threshold voltage $V_{th}$, the comparator generates the output spikes propagating to the next layer. During the firing, the reset MOSFET is also switched, allowing a current to follow from $T_2$ to $T_1$, thus driving the DW back to the initial position corresponding to the resetting of the neuron membrane voltage to its rest voltage.

## Multilayer Spintronic Leaky-Integrate and Fire Neuron Modeling

The LIF neurons are the fundamental units of the spiking neural networks, which involve thousands or millions of these units. To evaluate the circuit-level integration of the fabricated LIF neurons and proposed neurons. We developed compact models describing the LIF-characteristics in these devices. As discussed in the previous sections, the current pulses vs. measured output voltage show the typical integrate and leak behavior, which generates spikes in conjunction with the voltage comparator mimicking the LIF neuron. To model the characteristics of the measured devices, we first proposed the scaled multilayer device, as shown in the last section. The micromagnetic model complements the measured results, and LIF neuron characteristics are realized. The observed behavior in the estimated and micromagnetic devices is modeled based on the following physical concepts: (1) Magnetic Field-assisted current-driven DW motion. (2) DW pinning/depinning, (3) Demagnetization energy asymmetry caused by the DW motion, and (4) edge effects. Based on these forces, we propose the Multilayer Domain Wall LIF neuron model: The MTJ reading is used in the micromagnetic model to obtain the increased output voltage signal. The magnetization (m) of the free layer is considered the neuron's primary state variable (membrane potential). The magnetization depends on the DW As shown in Fig. 3(d-e), the current pulse shifts the DW, resulting in the free layer magnetization change directly proportional to the applied current density. When the current is off, the increasing demagnetization force and edge effects cause the pinning/depinning force DW to relax back. This leads to the magnetization leakage, which increases with the rising

DW position. The net magnetizationdepends upon the DW position X. We model the velocity of the DW, which computes the magnetization time evolution by:

$$\frac{dm}{dt} = -\alpha_D m^2 + \beta_D + \gamma_D J(t) \tag{4}$$

Where, $m$ is the membrane potential, threshold $= 0.75$, $\beta = -1e7$, $\alpha = -1.549$, $\gamma = 1.039e-3$, $J = 1e11$ and $dt = 1e-11$. Using the equation (12-13), and with some algebraic modifications, the magnetization dynamics is translated into Hall voltage and or MTJ output neuron dynamics as:

$$\frac{dV}{dt} = \frac{1}{\tau} V^2 + \beta_D + \gamma_D J(t) \tag{5}$$

The MTJ output voltage is discretized, and the reset condition is added to the model as:

$$V(t+1) = V(t) + dV$$

$$dV = \frac{\delta t}{\tau} \{ -V^2(t) + \frac{\alpha + \gamma J(t)}{\alpha \beta} \} \tag{6}$$

$$S = \{ 1, if \ V > V_{th}; 0 \ if \ V < V_{th}$$

The output spike S =1 also switches the reset transistor, allowing a short reset current pulse. Fig. 7(a) matches the measured LIF characteristics with the derived (proposed) LIF model, which qualitatively match each other to an accuracy above 90%. As shown the voltage increases sharply at the beginning, and with time, the leakage component starts dominating till the voltage reaches saturation; at this point, the integration and leakage almost cancel each other. The threshold voltage is set near the saturating point, and after that, the neuron is reset, as shown. Note: the voltage scale difference between the measured and model comes from the different reading mechanisms. The measurements ($\mu$Vs) were carried out as Hall measurements, whereas in the proposed model, we use MTJ reading to improve the signal; thus, the output is in the mV range.

Fig. 7(c) shows the ex-situ measured output spike frequency with increasing writing current for 6mA, 10mA, and 13mA, respectively; as shown in Fig. 7(c-d), we observe increasing spike rate with the increasing current from 6mA to 13mA. Similar results are obtained from the micromagnetic model-based LIF-MTJ neuron, as shown in Fig.7(e). The model's and measured devices' output spike rate improves with increasing current. These results and the discussion in Fig. 4 provide the benchmark for optimizing the range of writing current and biasing fields for tuning the LIF neuron response to the input stimuli.

## Spiking Neural Network Implementation

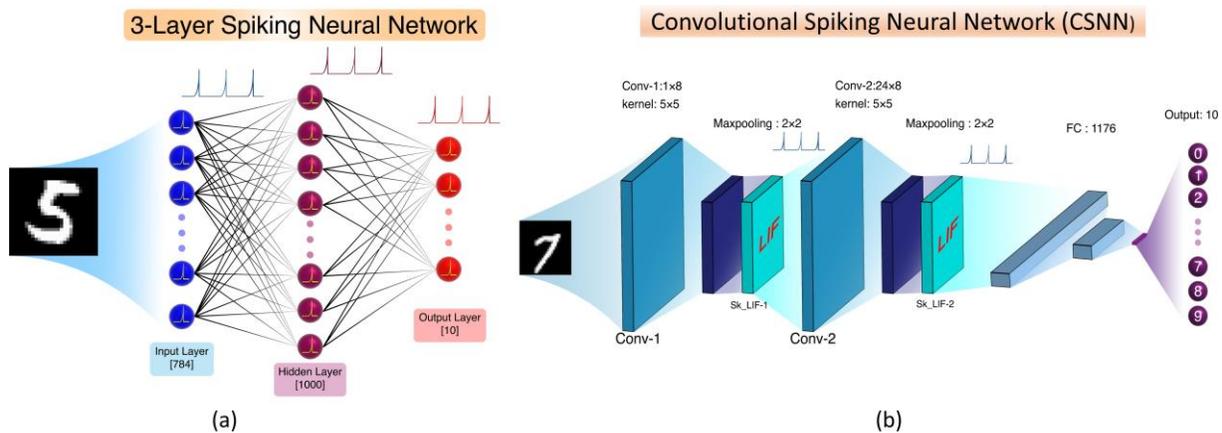

**Figure 8.** Neural network architectures:(a) 3-Layer FC-SNN with 784 input neurons, 1000 hidden neurons, and 10 output neurons. (b) Convolutional-SNN (CSNN) architecture -based on the proposed neuron models

We integrated the measured LIF devices and associated DW-LIF neuron models into two types of spiking neural network architectures to test the performance of the estimated LIF devices and the proposed neuron model. (1) The Fully Connected SNN and (See Fig. 8(a))(2) The Convolutional spiking neural network CSNN as shown in Fig. 8(b). We adopted the SNNTorch framework[36] Fig. 8(a) shows the 3-layer fully connected (FC) SNN architecture for the MNIST and Fashion MNIST data classification. The integration is carried out by using our models as the LIF neurons at each SNN node and CSNN node. The fully connected

SNN comprises 3- layers, with 784 input neurons, 1000 neurons in the hidden layer, and 10 neurons in the output layer. Conversely, the CSNN features two convolutional layers: Conv1 (1, 32, kernel=5, padding=5) and Conv2 (32, 64, kernel=5, padding=5), each followed by a MaxPool2d (kernel=2, stride=2) and the DW LIF neuron. At t=0, the membrane potential initializes to its rest value. During each time step, the input flattens to maintain correct dimensional integrity. It then passes to the layers, where input spikes increase the membrane potential and generate spikes on reaching the threshold. We recorded these dynamics in the spike and membrane tensors across all layers, repeating this process for 25 steps. The system then returns the spike and membrane recordings to the network and the CUDA device. As with the fully connected network, the input current and membrane potential pass through the network, and their dynamics are recorded over time.

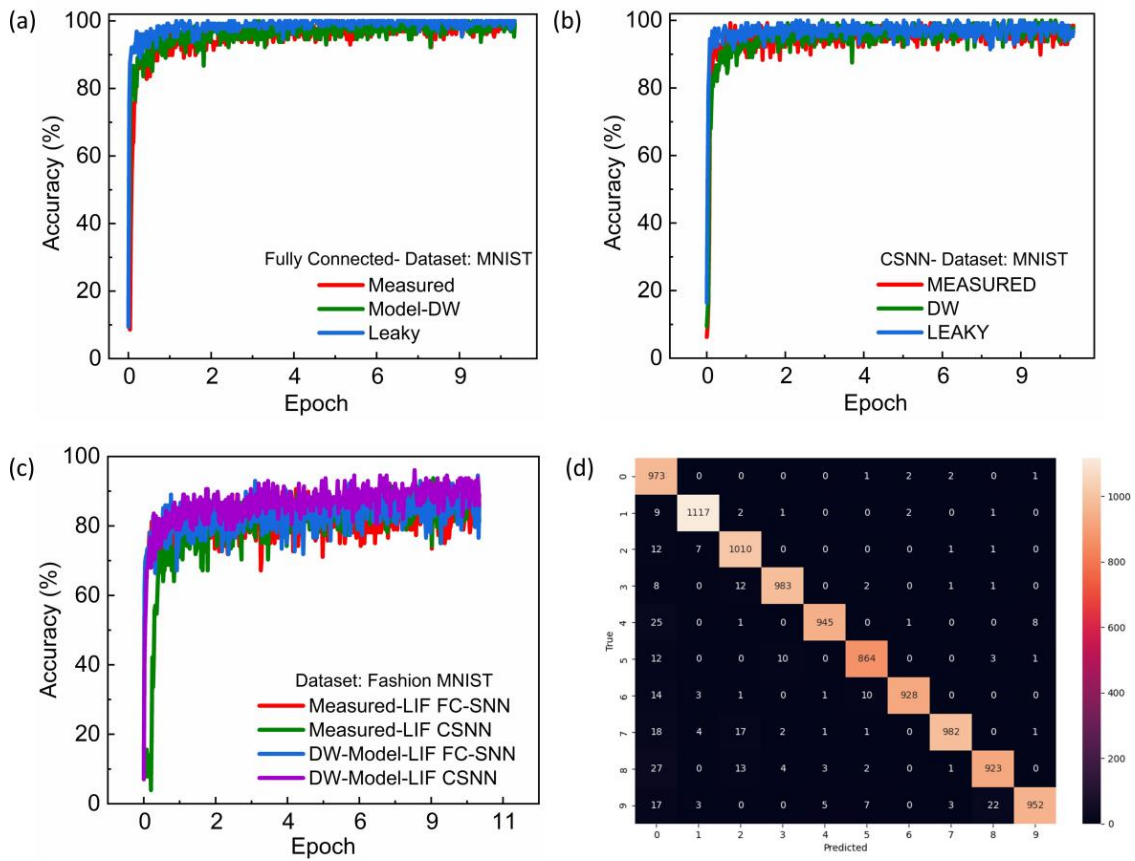

**Figure 9.** (a) Classification accuracy on MNIST dataset in FC-SNN architecture.
(b)Classification accuracy on MNIST dataset in convolutional CSNN architecture.
(c) Classification accuracy on Fashion-MNIST dataset in FC-SNN and CSNNarchitecture. (d) Confusion matrices

| Neuron Model | FC-MNIST | | CSNN-MNIST | | FC-FMNIST | | CSNN-FMNIST | |
|---|---|---|---|---|---|---|---|---|
| | Train Accuracy (%) | Test Accuracy (%) | Train Accuracy (%) | Test Accuracy (%) | Train Accuracy (%) | Test Accuracy (%) | Train Accuracy (%) | Test Accuracy (%) |
| DW-Model | 97.87 | 96.61 | 96.86 | 96.99 | 84.99 | 83.47 | 88.98 | 90.97 |
| Measured | 97.61 | 96.51 | 96.00 | 97.10 | 89 | 85 | 89 | 87.67 |
| SnnLeaky | 99.28 | 98.11 | 97.28 | 96.79 | NA | NA | NA | NA |

**Figure 10.** Table-2: Accuracy performance matrices of the LIF models.

We initialized variables within the neuron's class representing the membrane potential $\tau, \gamma, \beta, \alpha$ $V$ and $X$, $dt$, and the threshold values $V_t$ and $X_t$.

The neural network is trained by defining the Heaviside step function for the forward pass generating the spikes and the ArcTan function to facilitate gradient calculation during the backpropagation pass. The major problem faced in training SNNs is the non-differentiability of the spikes. This makes the computation of the gradient of the spike function challenging. The shifted Heaviside function is defined as:

$$S = \{ 1 \ v > v_{th}, \ 0 \ if \ v < v_{th} \}$$

The ArcTan function serves as a surrogate gradient, given by:

$$S \approx \frac{1}{\pi} \arctan(\pi v)$$

$$\frac{\partial S}{\partial v} = \frac{1}{\pi} \ \frac{1}{1 + (\pi v)^2}$$

(6)

Since spike values are binary (0s and 1s), their derivatives would inherently be zero, leading to the" dead neuron" problem. Researchers have proposed various smoothing functions to prevent this, including the sigmoid function, the fast sigmoid function, and the ArcTan functions. In our simulations, the ArcTan method proved superior.

Upon constructing the FC-SNN and CSNN, we designated the cross-entropy function for loss calculation and selected the Adam optimizer, achieving optimal results with a learning rate of 0.001.

The training loop, designed for ten epochs, includes an empty list for tracking loss history. The network iterates through the train loader, processing all data and respective targets, and collects final spike recordings. We initialize the loss tensor and compute loss based on spike summation and target values. The optimizer zero grad function clears previously calculated gradients, and loss val backward facilitates backpropagation. The optimizer step function updates the weights to optimize accuracy.

As shown in Fig. 9(a-d), the measured device LIF and the proposed DW-MTJ LIF models achieve a significant accuracy above 96% in all the MNIST dataset cases. Fig.9(a) shows the accuracy for 10 epochs for FC-SNN architecture on the MNIST dataset. The measured devices and corresponding models achieve training accuracy of around 97%, equivalent to the ideal snnLeaky model of the SNNTorch framework. Fig.9(b) shows the accuracy of the device models in a CSNN architecture and on the MNIST dataset. Here, the test accuracy is improved up to 97%, the same as the ideal snnLeaky model. To test the performance on the more complex dataset, in Fig. 9(c), the classification accuracy on the Fashion MNIST dataset is shown. The architectures remain unchanged, as SNN and CSNN in this case. As shown, we obtained an accuracy of around 91% on the FMNIST dataset.

Following training, we tested our devices and models with test data, analyzing accuracy with increasing iterations and epochs, and compared results with snnLeaky as a benchmark. The proposed neuron models achieved results comparable to the ideal SnnLeaky neuron model. The full details of the classification accuracy are summarised in Table 2. Fig. 9(d) shows one of the confusion matrices showing a correlation between the actual and predicted digits. These accuracy results, when added to the low latency neuron metrics and low writing energy requirements, underscore the viability of the proposed neuron models for large-scale energy-efficient neuromorphic computing applications.

# Methods

## Fabrication and Characterization

The Rotaris magnetron sputtering system deposited the magnetic thin films at room temperature on a 4-inch thermally oxidized Si wafer. A 300 nm bottom layer of $SiO_2$ is deposited before stack deposition by using thermal oxidation. Then, the multilayer stack is grown using ultrahigh vacuum magnetron sputtering at room temperature in $5 \times 10^{-8}$ mbar. The multilayer structure consists of, from the substrate side, a 300 nm $SiO_2$ insulating layer. Finally, a stack of $SiO_2$(300nm)/[Ta(5 nm)/CoFeB(x)/MgO(2 nm)] $\times$ 20 is formed, where MgO layers were deposited by RF magnetron sputtering, and the other layers were de- posited by DC magnetron sputtering. After the film deposition, we spin-coated AZ5214 photoresist with a thickness of 1.6 um and performed a hard bake for 2 mins at 110°C for positive tone use. We then patterned the crossbars on the resist using conventional photolithography. Ion beam etching using Ar gas removed the exposed magnetic stacks outside the resist mask. During the etching, we monitored the conductivity of the etched region and stopped the etching when the signal from the 300-nm-thick silicon oxide layer appeared. To form electrical contacts, Ti (10 nm)/Au (100 nm) was deposited on the sides of the crossbars through sputtering at a rate of 0.8 nm/s. The electrical contact pads were defined using photolithography and lifted off by immersion in acetone with ultrasonic processing for 5 mins.

## Characterization and Imaging

The magnetic characterization of the samples for thickness optimization was done using normal vibration sample magnetometry (VSM) at room temperature. After VSM, we performed imaging of the samples with multi-domain magnetic characteristics using magnetic force microscopy (MFM) based-Dimension Icon SPM. We used the CoIr-coated MFM tip provided by Bruker Inc. for probing. The electrical measurements were implemented in Quantum Design physical property measurement systems (PPMSs) Evercool II and Dynacool and a homemade probe station system equipped with an electromagnet.

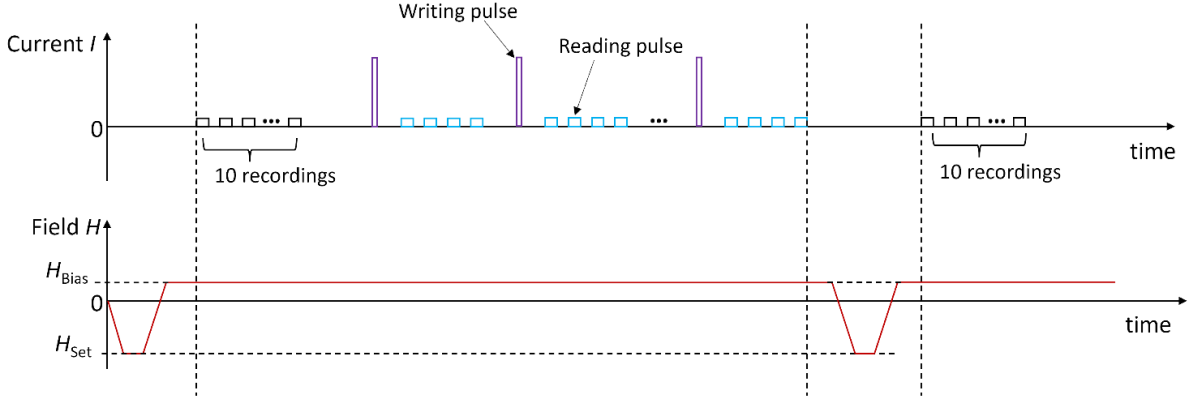

**Schematic diagram of the measurement procedure**

A Keithley model 6221 was used to source pulsed current and a Keithley model 2182A nanovoltmeter was used to measure the transverse voltage ($V_{xy}$). Two types of pulsed currents were utilized, i.e., writing and reading pulses, with pulse widths of 500 μs and 12 ms, respectively. Multiple amplitudes were used for the writing pulses, and 0.7 mA was used for the reading pulses. The transverse voltage was measured during the reading pulse, with a 2 ms delay relative to the start of the reading pulse. The measurement procedure for pulsed-current-induced transverse voltage change is as follows. Initially, a large out-of-plane magnetic field serving as a setting field (e.g., $H_{Set}$=-1800 Oe) was applied to achieve a saturated magnetization of the device. The magnetic field was then changed to a specific value defined as a bias field (e.g., $H_{Bias}$=+180 Oe) to produce a magnetic multidomain state of the device, followed by 10 transverse voltage recordings, each recording being the average of 3 individual measured voltages. Subsequently, a few writing pulses were injected to move magnetic domain walls and induce Hall voltage changes. After each writing pulse, 4 reading pulses were applied, and 4 transverse voltages were recorded. Finally, the magnetic field was changed to $H_{Set}$ to reset the magnetic state of the device, and it was then reduced again to $H_{Bias}$, followed by another 10 transverse voltage recordings. The timing diagram of the measurement procedure is shown below.

## Modelling and Simulations

The total magnetic energy of the free layer includes exchange, Zeeman, uniaxial anisotropy, demagnetization, and DMI energies.

$$\boldsymbol{E}(\boldsymbol{m}) = \int_{\boldsymbol{\nu}} \ [A(\nabla \boldsymbol{m})^2 - \mu_0 \boldsymbol{m} \cdot H_{\text{Ext}} - \frac{\mu_0}{2} \boldsymbol{m} \cdot H_d - \boldsymbol{K}_u(\hat{u} \cdot \boldsymbol{m}) + \varepsilon_{\text{DM}}]dv \tag{8}$$

where $A$ is the exchange stiffness, $\mu_0$ is the permeability, $K_u$ is the anisotropy energy density,

$H_d$ is the demagnetization field, and $H_{ext}$ is the external field; moreover, the DMI energy density is then computed as follows:

$$\varepsilon_{\text{DM}} = D\left[m_z(\nabla \cdot \boldsymbol{m}) - (\boldsymbol{m} \cdot \nabla) \cdot \boldsymbol{m}\right] \tag{9}$$

Solving the Euler equation yields the domain wall width ($\delta$) and the domain wall energy with DMI ($\sigma$), respectively

$$\delta = \pi\Delta = \pi\sqrt{\frac{\mathcal{A}}{\mathcal{K}_{eff}}} \tag{10}$$

$$\sigma = 4\sqrt{\mathcal{A}\mathcal{K}_{\text{eff}}} \mp \pi\text{D} \tag{11}$$

where $\Delta = A/K_{eff}$ is the Bloch wall width parameter.

Micromagnetic simulations were performed using MuMax[39,40] having the Landau–Lipschitz–Gilbert (LLG) equation as the basic magnetization dynamics computing unit. The LLG equation describes the magnetization evolution as follows:

$$\frac{d\hat{m}}{dt} = \frac{-\gamma}{1+\alpha^2} [\boldsymbol{m} \times \boldsymbol{H}_{eff} + \boldsymbol{m} \times (\boldsymbol{m} \times \boldsymbol{H}_{eff})] \tag{12}$$

where $\boldsymbol{m}$ is the normalized magnetization vector, $\gamma$ is the gyromagnetic ratio, $\alpha$ is the Gilbert damping coefficient, and

$$H_{eff} = \frac{-1}{\mu 0 MS} \frac{\delta \boldsymbol{E}}{\delta \boldsymbol{m}} \tag{13}$$

is the effective MF around which the magnetization process occurs. T The spin–orbit torque is then added in the form of modified STT in MuMax.[41]

$$\boldsymbol{\tau}_{SOT} = -\frac{\gamma}{1+\alpha^2} a_J [(1 + \xi\alpha)\boldsymbol{m} \times (\boldsymbol{m} \times \boldsymbol{p}) + (\xi - \alpha)(\boldsymbol{m} \times \boldsymbol{p})] \tag{14}$$

$$a_J = \left| \frac{\hbar \ \theta_{SH} j}{2M_S e} \right| \qquad \text{and} \qquad \boldsymbol{p} = \text{sign}(\theta_{SH}) \boldsymbol{j} \times \boldsymbol{n}$$

where $\theta_{SH}$ is the spin Hall coefficient of the material, $\boldsymbol{j}$ is the current density, and $d$ is the free layer thickness. The resistance of the proposed skyrmion MTJ synapse is then computed using the compact model presented in the main discussion Eqn. 2. We then consider the magnetization profile of the free layer and feed it to our model, which computes the resistance of the MTJ device as follows:

$$R_{syn} = \frac{V_{syn}}{I_{syn}} \tag{15}$$

# Conclusion

The spiking neural networks have gained much attention, considering their bio-plausibility and energy efficiency compared to ANNs. We presented the spintronic LIF neuron devices -based on the domain wall in the multilayer ferromagnetic structure. We showed how the combination of excitations, such as magnetic field, STT, SOT, thermal effects, and demagnetization energy, can be used to realize the leaky integration and fire LIF spintronic neurons. We showed the spintronic mem-transistor-like behavior, which provides dual tunability, thus versatility in device operation. The DW-MTJ LIF neurons are also proposed using the micromagnetic simulations showing very low spike latency, around 15ns, utilizing the writing energy of around 11.7pJ per spike. The proposed MTJ devices are scalable to 100x20nm, thus promising high-density data storage. The mathematical models of the measured and simulated devices that were developed have shown good integration capability in the FC- SNN and CSNN architectures. Achieving classification accuracy above 97% on the MNIST dataset and 90% on the FMNIST dataset. Considering all these essential device metrics, the proposed neurons and SNN schemes show an excellent prospect for spiking neural network applications and neuromorphic computing in general.

# Acknowledgments

# Author Contribution

A. H. Lone conceived the idea and fabricated the devices along with X. Zou. A. H. Lone and M. T. did the electrical characterization of the devices supported by D. Z and X. Z. A. L. did the micromagnetic simulations and neuron modeling, and D. N. R integrated the models in the SNN architecture and performed the simulations. A. H. Lone wrote the paper with support from D. N. R. and G. Setti. All other authors helped in the discussion of the results. G. Setti supervised the project.